\documentclass[12pt,a4paper]{article}

\usepackage[utf8]{inputenc}
\usepackage[T1]{fontenc}
\usepackage{lmodern}             
\input{glyphtounicode}
\usepackage[margin=2.5cm]{geometry}
\usepackage{graphicx}
\usepackage{booktabs}
\usepackage{longtable}
\usepackage{array}
\usepackage{hyperref}
\usepackage{natbib}
\usepackage{amsmath}

\usepackage{setspace}
\usepackage{fancyhdr}
\usepackage{xcolor}
\usepackage{listings}
\usepackage{caption}
\usepackage{subcaption}
\usepackage{float}

\hypersetup{
  colorlinks=true,
  linkcolor=blue!70!black,
  citecolor=blue!70!black,
  urlcolor=blue!70!black,
}

\begin{document}

\title{\textbf{A Multi-Survey Machine-Readable Corpus of Milky Way\\
Globular Cluster Parameters for Retrieval-Augmented\\
Generation Applications}}

\author{
  David C. Flynn\\
  \small EPS Research, Laurel, MD 20707, USA\\
  \small ORCID: 0000-0002-2768-6650\\
  \small \href{mailto:davidflynn@eps-research.com}{davidflynn@eps-research.com}
}

\date{Submitted to \textit{Publications of the Astronomical Society of the Pacific}}

\maketitle
\thispagestyle{fancy}

\begin{abstract}
We present the Milky Way Globular Cluster Corpus v1.3.1, a unified
machine-readable database of fundamental parameters for 174 Milky Way
globular clusters assembled from four independent published surveys. Each
cluster record integrates photometric, structural, and spectroscopically-calibrated
metallicity parameters from \citet{harris1996} (2010 revision), Gaia EDR3 proper motions from
\citet{vasiliev2021}, N-body dynamical masses and orbital parameters from
\citet{baumgardt2023}, and mean chemical abundances from the APOGEE DR17
globular cluster Value Added Catalog of \citet{schiavon2024}. The corpus
contains 17,438 non-null data points across 174 clusters stored in JSONL,
JSON, and flat CSV formats with consistent native-typed fields
(\texttt{float}, \texttt{int}, \texttt{bool}, \texttt{null}), embedded
provenance blocks, and fully documented schema. Survey coverage is 157/174
clusters for Harris photometry, 170/174 for Gaia EDR3 proper motions,
154/174 for Baumgardt N-body dynamics, and 72/174 for APOGEE DR17
chemistry. The corpus was designed as a Retrieval-Augmented Generation
(RAG) knowledge base for large language model applications in astrophysics
research, following the same multi-survey integration methodology as the
Unified Galaxy HI Rotation Curve Corpus \citep{flynn2026sparc}, and has
been validated for structured context injection with instruction-following
language models. It is equally suitable for traditional quantitative
analyses including orbit modeling, cluster classification, chemical
tagging, and multi-survey cross-validation. The dataset is available at
Zenodo DOI: \href{https://doi.org/10.5281/zenodo.19907766}{10.5281/zenodo.19907766}.
\end{abstract}

\noindent\textbf{Keywords:} globular clusters --- Milky Way --- stellar kinematics ---
chemical abundances --- N-body models --- proper motions --- Gaia EDR3 ---
APOGEE --- catalogs --- databases --- methods: data analysis

\section{Introduction}
\label{sec:intro}

The Milky Way globular cluster system is one of the most thoroughly studied
populations in astrophysics. These old, dense stellar systems serve as
tracers of Galactic structure and chemical evolution, probes of dark matter
distribution, and benchmarks for stellar population models. Decades of
multi-wavelength observational campaigns have produced a rich but fragmented
literature: photometric catalogs, N-body dynamical models, Gaia astrometry,
and high-resolution spectroscopic surveys exist as independent publications,
each with its own format, naming convention, and coverage.

The rise of large language models (LLMs) and Retrieval-Augmented Generation
(RAG) architectures in scientific research creates a new demand:
machine-readable corpora that are not merely tabular databases but structured
knowledge representations accessible to both programmatic analysis and
natural language inference pipelines. A RAG corpus must be consistently
typed, null-safe, self-describing, and organized so that each record is a
semantically coherent unit --- a single cluster carrying all available
information about that object from all available sources.

This paper describes the construction and content of the Milky Way Globular
Cluster Corpus v1.3.1, which satisfies these requirements. The corpus follows
the design principles established in the Unified Galaxy HI Rotation Curve
Corpus \citep{flynn2026sparc}, extending the multi-survey integration approach
from galaxy kinematics to globular cluster physics.

The four source surveys were chosen to provide orthogonal physical
information: \citet{harris1996} for photometry and classical structure;
\citet{vasiliev2021} for space astrometry; \citet{baumgardt2023} for
dynamical modeling; and \citet{schiavon2024} for high-resolution
spectroscopic chemistry. No single survey covers all 174 clusters in the
corpus; coverage fractions range from 41\% (APOGEE DR17) to 98\%
(Vasiliev \& Baumgardt 2021), with the Harris catalog forming the
157-cluster backbone.

\section{Source Surveys}
\label{sec:surveys}

\subsection{Harris (1996, 2010 Edition)}
\label{sec:harris}

The Harris catalog \citep{harris1996}, revised 2010, is the standard reference
compilation for Milky Way GC parameters, providing photometric and structural
data for 157 clusters. The spectroscopic metallicity $[\mathrm{Fe/H}]$
(Carretta et al.\ scale) entries in the Harris catalog are derived from the
homogeneous metallicity scale of \citet{carretta2009}, based on high-resolution
($R \sim 45{,}000$) UVES optical spectroscopy; they are not photometric
estimates. Parameters include: positions in equatorial and
Galactic coordinates; heliocentric distance $R_\odot$ and Galactocentric
distance $R_\mathrm{GC}$; Galactocentric $X/Y/Z$ coordinates; reddening
$E(B-V)$; spectroscopic metallicity $[\mathrm{Fe/H}]$ (Carretta et al.\ scale); apparent and absolute $V$
magnitudes; distance modulus; horizontal branch magnitude $V_\mathrm{HB}$;
color indices $U-B$, $B-V$, $V-R$, $V-I$; ellipticity; spectral type; King
concentration parameter $c$ \citep{king1966}; core-collapse classification;
core and half-light radii in arcmin and kpc; central surface brightness
$\mu_V$; central logarithmic density $\log\rho_0$; core and half-mass
relaxation times; mean radial velocity and velocity dispersion.

\subsection{Vasiliev \& Baumgardt (2021)}
\label{sec:vasiliev}

\citet{vasiliev2021} published a comprehensive Gaia EDR3 astrometric analysis
of 170 Milky Way GCs. Their mixture model analysis yields mean proper motions
$\mu_{\alpha*}$ and $\mu_\delta$ with errors and correlation coefficient, mean
Gaia parallax with \citet{lindegren2021} zero-point correction applied
per-star, a Plummer scale radius, and the number of member stars with good
astrometry.

This survey provides 17 clusters not present in the Harris 2010 catalog,
representing systems discovered after the Harris compilation. Four clusters
(2MS-GC01, 2MS-GC02, GLIMPSE01, GLIMPSE02) lie behind extreme dust columns
($E(B-V)$ ranging from 5.2 to 34.5) that Gaia optical astrometry cannot
penetrate; their \texttt{gaia\_edr3} blocks are null throughout.

\subsection{Baumgardt et al.\ (2023), v4 N-body Database}
\label{sec:baumgardt}

The Baumgardt et al.\ N-body database \citep[v4, March 2023;][]{baumgardt2018,
baumgardt_vasiliev2021, baumgardt2023} provides parameters derived by fitting
multimass N-body models to compilations of ground-based radial velocities,
Gaia DR3 proper motions, and HST-based stellar mass functions. The database
covers 154 clusters in the current corpus.

Two sub-tables were ingested. The orbits table provides precise distances with
errors, mean radial velocity, Gaia DR3 proper motions, Galactocentric $X/Y/Z$
positions and $U/V/W$ space velocities with errors, and orbital pericenter
$r_\mathrm{peri}$ and apocenter $r_\mathrm{apo}$ computed in the
\citet{irrgang2013} Galactic potential. The structural parameters table
provides total dynamical mass with uncertainty, $V$-band magnitude and $M/L$
ratio, core radius $r_c$, projected half-light radius $r_\mathrm{hl}$, 3D
half-mass radius $r_\mathrm{hm}$, and tidal radius $r_t$ --- all in parsecs
--- central and half-mass logarithmic densities, half-mass relaxation time,
initial mass, dissolution timescale, global IMF slope $\alpha$ (Salpeter
$= -2.3$), central 1D velocity dispersion $\sigma_0$, central escape
velocity, mass segregation parameters $\eta_c$ and $\eta_h$
\citep{trenti2013}, and rotation amplitude $A_\mathrm{rot}$ with detection
probability $P_\mathrm{rot}$ \citep{sollima2019}.

Five clusters present in the Baumgardt database but not in Harris or Vasiliev
are appended: Gran 2, Gran 3, Gran 5, Patchick 126, and VVV-CL160.

\subsection{Schiavon et al.\ (2024), APOGEE DR17 GC Value Added Catalog}
\label{sec:apogee}

\citet{schiavon2024} constructed the APOGEE DR17 GC Value Added Catalog by
identifying member stars in 72 Milky Way GCs from the APOGEE DR17 allStar
catalog \citep{abdurrouf2022}, yielding 6,422 member star entries. APOGEE
acquires high-resolution ($R \sim 22{,}500$) $H$-band spectra using the
300-fiber APOGEE spectrographs. Stellar parameters and abundances were derived
using the ASPCAP pipeline \citep{garciaperez2016} with the DR17
\texttt{synspec\_rev1} synthetic grid incorporating NLTE treatment for Na,
Mg, K, and Ca.

From this VAC we ingest cluster-level mean values: mean $[\mathrm{Fe/H}]$,
mean radial velocity with error, heliocentric and Galactocentric distances
from \citet{baumgardt_vasiliev2021}, cluster mass in units of $10^4\,M_\odot$,
Jacobi radius in degrees, and total number of member star entries in the VAC.
The 72-cluster coverage (41\%) reflects APOGEE fiber allocation priorities and
$H$-band extinction limits.

\section{Data Integration Methodology}
\label{sec:method}

\subsection{Cluster Identification and Name Resolution}

Primary cluster identifiers follow the Harris 2010 naming convention where
available. Baumgardt and Vasiliev catalog names were mapped to Harris primary
identifiers through an explicit lookup table, verified against the SIMBAD
astronomical database.

\subsection{Data Type Standardization}

All numeric values are stored as native Python \texttt{float} or \texttt{int}
primitives. Missing or unmeasured values are stored as JSON \texttt{null}.
Errors are stored as separate \texttt{\_err} fields. Boolean classifications
(\texttt{core\_collapsed}, \texttt{inner\_galaxy}, \texttt{sgr\_stream}) are
native JSON booleans. No unit conversions were performed without documentation.
Radii in arcmin (Harris) and parsecs (Baumgardt) are both preserved in the
schema because the two sets of values are derived from different methods and
should be kept distinct.

\subsection{Null Handling and Coverage Gaps}

Null values arise from three distinct causes. \textit{Physical
inaccessibility}: the four Gaia-invisible clusters have null Gaia blocks
because their dust columns prevent optical astrometry. \textit{Survey coverage
limits}: APOGEE DR17 observed 72 of 174 clusters; Baumgardt v4 covers 154 of
174; uncovered clusters lack sufficient member stars for N-body fitting or
fell outside observed fields. \textit{Catalog vintage}: the 22 clusters not in
Harris 2010 have null Harris fields because they were discovered after the 2010
revision. All three causes are documented in the provenance blocks embedded in
each cluster record.

\subsection{Provenance Embedding}

Each data block carries a provenance sub-object containing the source
citation, DOI or URL, and methodological notes. Provenance is embedded at the
block level to balance completeness with compactness, ensuring that any record
retrieved from the JSONL corpus carries its own attribution.

\subsection{Known Issues and Bug Fix History}

\textbf{v1.3.1 patch.} Three bugs identified during peer review were corrected
before the Zenodo deposit. \textit{Bug 1}: 14 clusters with two-word alt-names
(NGC 104/47~Tuc, NGC 1904/M~79, NGC 4590/M~68, NGC 5024/M~53, NGC 5272/M~3,
NGC 5904/M~5, NGC 6205/M~13, NGC 6218/M~12, NGC 6093/M~80, NGC 6266/M~62,
NGC 6273/M~19, NGC 6656/M~22, NGC 6779/M~56, Ton~2) had a one-column
rightward shift in $l$, $b$, and Harris distance fields. The
\texttt{baumgardt2023} positional fields serve as the authoritative source for
positional queries for all 154 covered clusters. \textit{Bug 2}: 2MS-GC01 and
GLIMPSE01 had incorrect \texttt{feh}, \texttt{feh\_weight}, and \texttt{ebv}
values due to a zero-weight parse error. \textit{Bug 3}: the horizontal branch
magnitude \texttt{v\_hb} was absent from the flat CSV export. All three are
fixed in v1.3.1.

\subsubsection*{Note on the \texttt{inner\_galaxy} flag}
\label{sec:flags-note}
The \texttt{inner\_galaxy} boolean is inherited from an upstream
classification list and is \textit{not} computed from the current
\texttt{baumgardt2023.r\_gc\_kpc} values in this corpus. Cross-checking the
flag against the Baumgardt (2023) Galactocentric distances reveals
inconsistencies that arise because the upstream flag was assigned using
older distance estimates (predominantly Harris 2010) that have since been
superseded. For example, UKS~1 carries \texttt{inner\_galaxy = true}
based on its Harris distance of $R_\mathrm{GC} = 0.7$ kpc, but the
Baumgardt (2023) N-body fit places it at $R_\mathrm{GC} = 7.47$ kpc; the
flag was not updated when the new distance was ingested. NGC~104
(47~Tuc) likewise carries the flag set to \texttt{true} despite a
canonical halo/thick-disk location ($l=306^\circ$, $b=-45^\circ$,
$R_\mathrm{GC} = 7.5$ kpc), reflecting an artifact of the upstream list
rather than a current chemodynamical assignment. Conversely, NGC~6553 at
$R_\mathrm{GC} = 2.37$ kpc is flagged \texttt{false}.
Users requiring a clean radial cut should filter directly on
\texttt{baumgardt2023.r\_gc\_kpc} rather than relying on the flag. The
flag is retained in v1.3.1 for backward compatibility with consumers of
the upstream classification; a forthcoming v1.4 release will recompute
\texttt{inner\_galaxy} from the Baumgardt distances and document the
threshold explicitly.

\section{Schema Description}
\label{sec:schema}

Table~\ref{tab:schema} gives the complete per-cluster JSON schema. The flat
CSV exposes all fields at one level using block-prefix naming: no prefix for
Harris/identity fields, \texttt{b\_} for Baumgardt 2023, \texttt{gaia\_} for
Vasiliev 2021, \texttt{a\_} for APOGEE DR17.

\begin{longtable}{p{2.6cm}p{4.2cm}p{1.8cm}p{5.5cm}}
\caption{Per-cluster JSON schema for the Milky Way Globular Cluster Corpus v1.3.1.}
\label{tab:schema}\\
\toprule
\textbf{Block} & \textbf{Field(s)} & \textbf{Type} & \textbf{Description} \\
\midrule
\endfirsthead
\multicolumn{4}{c}{\tablename\ \thetable{} -- continued}\\
\toprule
\textbf{Block} & \textbf{Field(s)} & \textbf{Type} & \textbf{Description} \\
\midrule
\endhead
\bottomrule
\endfoot
\bottomrule
\endlastfoot
(top)         & \texttt{cluster\_id}          & str           & Primary identifier \\
(top)         & \texttt{alt\_name}            & str$|$null    & Common name \\
position      & \texttt{ra\_hms, dec\_dms}    & str$|$null    & Equatorial coords \\
position      & \texttt{l\_deg, b\_deg}       & float$|$null  & Galactic $l$, $b$ (deg) \\
distances     & \texttt{r\_sun\_kpc}          & float$|$null  & Heliocentric distance (kpc) \\
distances     & \texttt{r\_gc\_kpc}           & float$|$null  & Galactocentric distance (kpc) \\
distances     & \texttt{x/y/z\_kpc}           & float$|$null  & Galactocentric $X/Y/Z$ (kpc) \\
metallicity   & \texttt{feh}                  & float$|$null  & $[\mathrm{Fe/H}]$ (Harris) \\
metallicity   & \texttt{feh\_weight}          & int$|$null    & Metallicity weight \\
metallicity   & \texttt{ebv}                  & float$|$null  & Reddening $E(B-V)$ \\
photometry    & \texttt{v\_hb, dist\_mod}     & float$|$null  & $V_\mathrm{HB}$, distance modulus \\
photometry    & \texttt{v\_t, m\_v\_t}        & float$|$null  & Apparent/absolute $V$ mag \\
photometry    & \texttt{ellipticity}          & float$|$null  & Ellipticity \\
photometry    & \texttt{colors \{ub,bv,vr,vi\}} & float$|$null & Color indices \\
kinematics    & \texttt{v\_r\_kms $\pm$ err}  & float$|$null  & Heliocentric RV (km\,s$^{-1}$) \\
kinematics    & \texttt{sig\_v\_kms $\pm$ err}& float$|$null  & Velocity dispersion (km\,s$^{-1}$) \\
structure     & \texttt{king\_concentration}  & float$|$null  & King concentration $c$ \\
structure     & \texttt{core\_collapsed}      & bool          & Core-collapse flag \\
structure     & \texttt{r\_core/half\_arcmin} & float$|$null  & Core/half-light radii (arcmin) \\
structure     & \texttt{r\_core/half\_kpc}    & float$|$null  & Core/half-light radii (kpc) \\
structure     & \texttt{mu\_v\_central, log\_rho0} & float$|$null & Central brightness/density \\
dynamics      & \texttt{log\_t\_rc/rh\_yr}    & float$|$null  & Relaxation times ($\log$ yr) \\
flags         & \texttt{inner\_galaxy}        & bool          & Inner-Galaxy association flag (legacy; see Sec.~\ref{sec:flags-note}) \\
flags         & \texttt{sgr\_stream}          & bool          & Sgr dSph association \\
gaia\_edr3    & \texttt{mu\_alpha\_mas\_yr $\pm$ err} & float$|$null & PM in $\alpha\cos\delta$ (mas\,yr$^{-1}$) \\
gaia\_edr3    & \texttt{mu\_delta\_mas\_yr $\pm$ err} & float$|$null & PM in $\delta$ (mas\,yr$^{-1}$) \\
gaia\_edr3    & \texttt{corr\_mu}             & float$|$null  & PM correlation coefficient \\
gaia\_edr3    & \texttt{parallax\_mas $\pm$ err} & float$|$null & Gaia parallax (mas) \\
gaia\_edr3    & \texttt{n\_members\_gaia}     & int$|$null    & $N$ stars with good astrometry \\
baumgardt2023 & \texttt{r\_sun/gc\_kpc $\pm$ err} & float$|$null & Distances (kpc) \\
baumgardt2023 & \texttt{rv\_kms $\pm$ err}    & float$|$null  & Mean RV (km\,s$^{-1}$) \\
baumgardt2023 & \texttt{x/y/z\_kpc $\pm$ err} & float$|$null & Galactocentric position (kpc) \\
baumgardt2023 & \texttt{u/v/w\_kms $\pm$ err} & float$|$null & Space velocities (km\,s$^{-1}$) \\
baumgardt2023 & \texttt{r\_peri/apo\_kpc $\pm$ err} & float$|$null & Orbital pericenter/apocenter (kpc) \\
baumgardt2023 & \texttt{n\_rv, n\_pm}         & int$|$null    & $N$ stars with RV/PM \\
baumgardt2023 & \texttt{mass\_msun $\pm$ err} & float$|$null  & Dynamical mass ($M_\odot$) \\
baumgardt2023 & \texttt{rc/rhl/rhm/rt\_pc}    & float$|$null  & Structural radii (pc) \\
baumgardt2023 & \texttt{sigma0\_kms, v\_esc\_kms} & float$|$null & Central dispersion, $v_\mathrm{esc}$ \\
baumgardt2023 & \texttt{mf\_slope $\pm$ err}  & float$|$null  & Global IMF slope $\alpha$ \\
baumgardt2023 & \texttt{eta\_c, eta\_h}       & float$|$null  & Mass segregation parameters \\
baumgardt2023 & \texttt{a\_rot\_kms $\pm$ err, p\_rot\_pct} & float$|$null & Rotation amplitude/probability \\
apogee\_dr17  & \texttt{feh\_apogee}          & float$|$null  & Mean $[\mathrm{Fe/H}]$ (APOGEE) \\
apogee\_dr17  & \texttt{rv\_mean\_kms $\pm$ err} & float$|$null & Mean RV (km\,s$^{-1}$) \\
apogee\_dr17  & \texttt{mass\_1e4\_msun}      & float$|$null  & Mass ($10^4\,M_\odot$) \\
apogee\_dr17  & \texttt{r\_jacobi\_deg}       & float$|$null  & Jacobi radius (deg) \\
apogee\_dr17  & \texttt{n\_members}           & int$|$null    & $N$ member stars in VAC \\
\end{longtable}

\section{Example Records}
\label{sec:examples}

The following three records illustrate the range of data coverage. Provenance
sub-blocks are omitted for brevity.

\subsection{NGC 104 (47 Tuc) --- Full Four-Survey Record}

NGC 104 is the most data-rich record in the corpus. All four source surveys
contribute: Harris provides the foundational photometry; \citet{vasiliev2021}
yield 39,932 Gaia member stars ($\sigma_\mu \approx 0.008$ mas\,yr$^{-1}$);
\citet{baumgardt2023} derive a total mass of $853{,}000\,M_\odot$ with
$\sigma_0 = 11.9$ km\,s$^{-1}$ and an essentially circular orbit
($r_\mathrm{peri} = 5.47$ kpc, $r_\mathrm{apo} = 7.51$ kpc); and
\citet{schiavon2024} measure a mean $[\mathrm{Fe/H}] = -0.74$ from 297 member
giants, consistent with the Harris spectroscopic value of $-0.72$ (Carretta et al.\ scale).

\begin{lstlisting}[language={}]
{"cluster_id": "NGC 104", "alt_name": "47 Tuc",
 "position": {"l_deg": 305.89, "b_deg": -44.89},
 "metallicity": {"feh": -0.72, "ebv": 0.04},
 "structure": {"king_concentration": 2.07, "core_collapsed": false,
               "r_core_kpc": 0.032, "r_half_kpc": 0.2821},
 "gaia_edr3": {"mu_alpha_mas_yr": 5.252, "mu_alpha_err": 0.021,
               "mu_delta_mas_yr": -2.551, "n_members_gaia": 39932},
 "baumgardt2023": {"mass_msun": 853000.0, "rc_pc": 0.61,
                   "sigma0_kms": 11.9, "r_peri_kpc": 5.47,
                   "r_apo_kpc": 7.51, "mf_slope": -0.65},
 "apogee_dr17": {"feh_apogee": -0.74, "n_members": 297}}
\end{lstlisting}

\subsection{2MS-GC01 --- Gaia-Invisible Record}

2MS-GC01 was discovered through infrared photometry. Its extreme reddening
($E(B-V) = 6.80$) renders it invisible to Gaia optical detectors. Harris
provides limited photometry and a radial velocity from near-infrared
spectroscopy. The \texttt{gaia\_edr3}, \texttt{baumgardt2023}, and
\texttt{apogee\_dr17} blocks are null throughout, representing physical
inaccessibility rather than a survey coverage gap.

\begin{lstlisting}[language={}]
{"cluster_id": "2MS-GC01", "alt_name": "2MASS-GC01",
 "position": {"l_deg": 10.48, "b_deg": 0.11},
 "metallicity": {"feh": null, "feh_weight": 0, "ebv": 6.80},
 "kinematics": {"v_r_kms": 0.85, "sig_v_kms": 8.43},
 "gaia_edr3": null,
 "baumgardt2023": null,
 "apogee_dr17": null}
\end{lstlisting}

\subsection{Bliss 1 --- Post-Harris Discovery, Gaia-Only Record}

Bliss 1 was discovered after the Harris 2010 revision. All Harris blocks are
null. The Gaia EDR3 block provides a proper motion solution from 10 member
stars. Baumgardt and APOGEE blocks are null because 10 members are
insufficient for N-body fitting. Galactic coordinates were computed from
the Gaia RA/Dec using \textsc{astropy} \citep{astropy2022}.

\begin{lstlisting}[language={}]
{"cluster_id": "Bliss 1", "alt_name": null,
 "position": {"l_deg": 290.8321, "b_deg": 19.6528},
 "metallicity": {"feh": null, "ebv": null},
 "gaia_edr3": {"ra_deg": 177.511, "dec_deg": -41.772,
               "mu_alpha_mas_yr": -2.34, "mu_alpha_err": 0.042,
               "mu_delta_mas_yr": 0.138, "mu_delta_err": 0.038,
               "n_members_gaia": 10},
 "baumgardt2023": null,
 "apogee_dr17": null}
\end{lstlisting}

\section{Coverage Analysis}
\label{sec:coverage}

\begin{figure}[H]
\centering
\includegraphics[width=\textwidth]{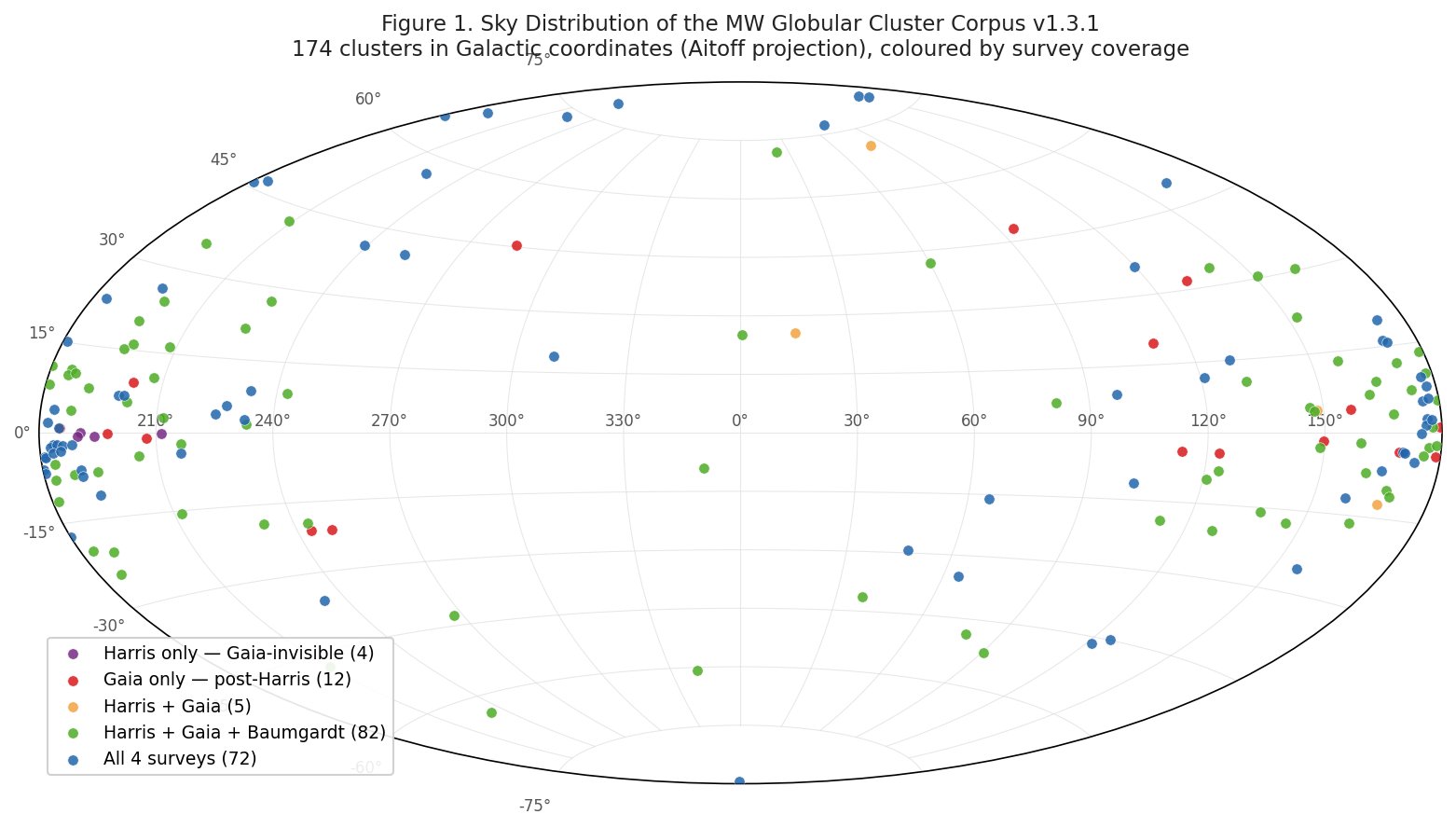}
\caption{Sky distribution of the 174 clusters in the Milky Way Globular
Cluster Corpus v1.3.1, shown in Galactic coordinates (Aitoff projection) and
coloured by survey coverage tier. Blue: all four surveys (72 clusters). Green:
Harris + Gaia + Baumgardt (82). Orange: Harris + Gaia only (5). Red: Gaia
only, post-Harris discovery (12). Purple: Harris only, Gaia-invisible (4). The
strong concentration near $l = 0^\circ$ reflects the Galactic bulge and inner
halo GC population.}
\label{fig:sky}
\end{figure}

\begin{figure}[H]
\centering
\includegraphics[width=0.8\textwidth]{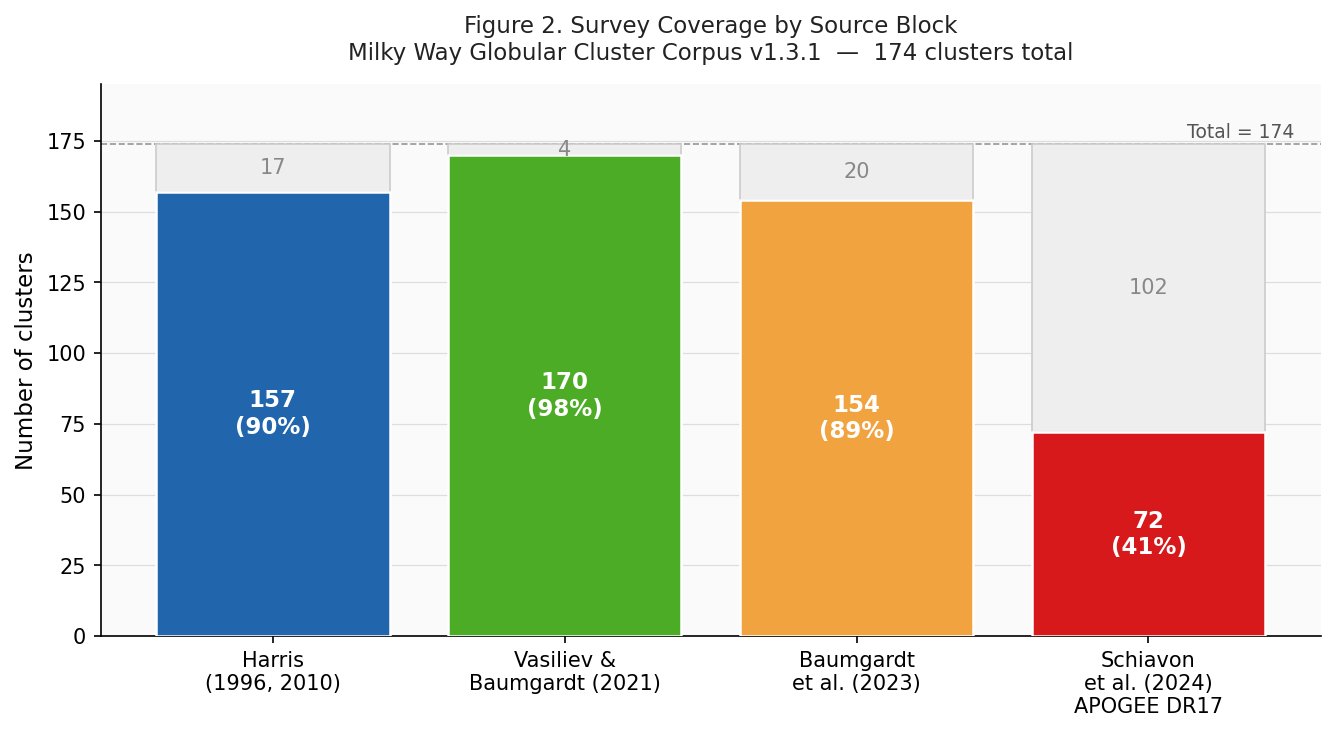}
\caption{Survey coverage by source block. Each bar shows the number and
percentage of clusters with non-null data in that block. Grey portions
indicate uncovered clusters; the reason for each gap differs by block (see
Table~\ref{tab:coverage} and Section~\ref{sec:method}).}
\label{fig:coverage}
\end{figure}

Figure~\ref{fig:sky} shows the sky distribution of all 174 clusters in
Galactic coordinates, coloured by survey coverage tier. The strong
concentration toward the Galactic centre ($l \sim 0^\circ$, $b \sim 0^\circ$)
reflects the underlying distribution of the MW GC system, with APOGEE
coverage (blue) following the same trend since APOGEE targets were
preferentially drawn from accessible, non-reddened fields.
Figure~\ref{fig:coverage} summarises the per-block coverage fractions.

Table~\ref{tab:coverage} gives the coverage summary. The 12 clusters with
Gaia PM data only are among the most recently discovered systems and will
accumulate additional parameters as follow-up observations are published.

\begin{table}[h]
\centering
\caption{Coverage summary by source block.}
\label{tab:coverage}
\begin{tabular}{llrl}
\toprule
\textbf{Source} & \textbf{Clusters} & \textbf{Fraction} & \textbf{Primary gap reason} \\
\midrule
Harris (1996, 2010 ed.)         & 157/174 & 90\% & 17 post-2010 discoveries \\
Vasiliev \& Baumgardt (2021)    & 170/174 & 98\% & 4 Gaia-invisible (extreme extinction) \\
Baumgardt et al.\ (2023)        & 154/174 & 89\% & Insufficient members for N-body \\
Schiavon et al.\ (2024) APOGEE  &  72/174 & 41\% & Fiber coverage + extinction limits \\
\midrule
\textbf{Total}                  & \textbf{174} & & \textbf{17,438 non-null data points} \\
\bottomrule
\end{tabular}
\end{table}

\section{Scientific Figures}
\label{sec:figures}

\begin{figure}[H]
\centering
\includegraphics[width=\textwidth]{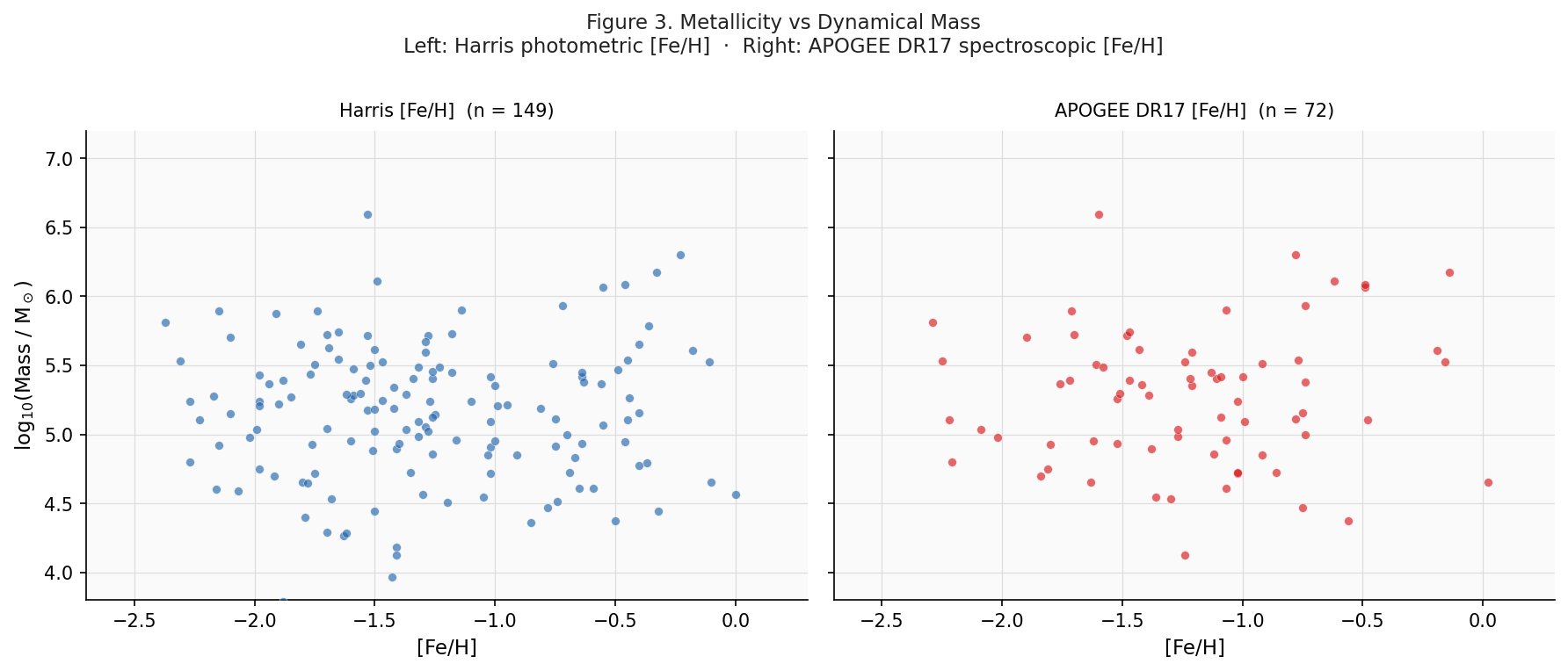}
\caption{Metallicity vs.\ dynamical mass for the Milky Way globular cluster
corpus. Left: Harris (1996, 2010) spectroscopic $[\mathrm{Fe/H}]$ (Carretta et al.\ scale)
vs.\
Baumgardt et al.\ (2023) $\log_{10}$ dynamical mass ($n = 149$ clusters).
Right: APOGEE DR17 spectroscopic $[\mathrm{Fe/H}]$ vs.\ the same mass
($n = 72$). Both panels represent independent high-resolution spectroscopic
pipelines (UVES optical at $R \sim 45{,}000$ vs.\ APOGEE H-band at
$R \sim 22{,}500$); their agreement demonstrates cross-pipeline spectroscopic
consistency within the corpus. The broad scatter and lack of strong
correlation is consistent with the known independence of GC mass from
metallicity across the full population.}
\label{fig:fehmass}
\end{figure}

Figure~\ref{fig:fehmass} shows the relationship between cluster metallicity
and dynamical mass for both the Harris spectroscopic $[\mathrm{Fe/H}]$
(Carretta et al.\ scale; left, $n = 149$) and the APOGEE DR17 spectroscopic
$[\mathrm{Fe/H}]$ (right, $n = 72$). The broad scatter with no strong
correlation is consistent with the known independence of GC mass and
metallicity across the full Galactic GC population. Both panels use
high-resolution spectroscopic $[\mathrm{Fe/H}]$ --- the Harris values
calibrated to the \citet{carretta2009} UVES scale ($R \sim 45{,}000$,
optical), and APOGEE DR17 from the ASPCAP pipeline ($R \sim 22{,}500$,
H-band) --- so their agreement demonstrates cross-pipeline spectroscopic
consistency across two independent high-resolution surveys.

\begin{figure}[H]
\centering
\includegraphics[width=0.75\textwidth]{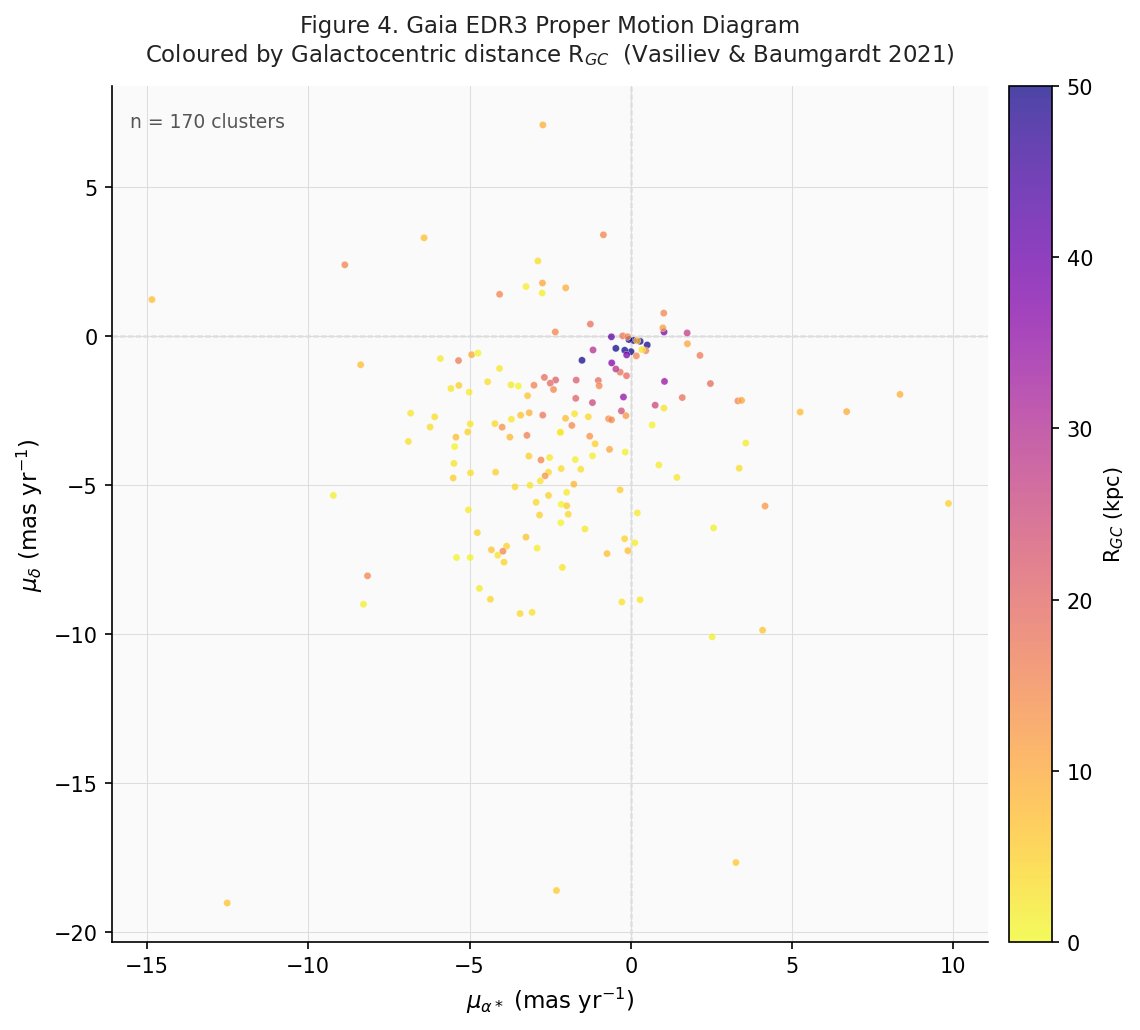}
\caption{Gaia EDR3 proper motion diagram ($\mu_{\alpha*}$ vs.\ $\mu_\delta$)
for 170 clusters with Vasiliev \& Baumgardt (2021) measurements, coloured by
Galactocentric distance $R_\mathrm{GC}$ from the Baumgardt et al.\ (2023)
database. The concentration near zero reflects the bulk Galactic rotation
frame, with outer-halo clusters (blue/purple) exhibiting larger proper motion
amplitudes from more eccentric orbits.}
\label{fig:pm}
\end{figure}

Figure~\ref{fig:pm} shows the Gaia EDR3 proper motion diagram for all 170
clusters with PM measurements, coloured by Galactocentric distance $R_\mathrm{GC}$.
The concentration of inner-halo clusters (yellow/orange) near the origin
reflects the near-zero net PM expected for an isotropic, pressure-supported
population at small $R_\mathrm{GC}$, while outer-halo clusters (blue/purple)
show the larger PM amplitudes associated with more radial orbits.

\section{Use Cases}
\label{sec:usecases}

The corpus is designed to support a broad range of scientific and
computational applications.

\textbf{RAG and LLM applications.} The JSONL format allows each cluster record
to be retrieved as a semantically complete unit by a vector database or
retrieval pipeline. Embedded provenance blocks ensure retrieved facts carry
their citations, supporting verifiable AI-assisted literature synthesis. The
deterministic JSONL structure stabilizes the token context window for
inference; the corpus has been validated against ten instruction-following
LLM systems spanning frontier cloud models (Claude Opus~4.6, Claude
Haiku~4.5, Microsoft Copilot~Pro, Google Gemini~Pro) and local open-weight
models from 1.5{B} to 70{B} parameters, and is suitable for deployment with
any instruction-following model supporting structured context injection.

\textbf{Cluster classification and machine learning.} The flat CSV provides
a 82-column feature matrix suitable for use in scikit-learn, PyTorch, or
R-based classifiers. Natural targets include core-collapse classification
(29 known cases, boolean label), Sagittarius stream membership (6 clusters),
and inner-Galaxy vs.\ halo membership.

\textbf{Orbit modeling and dynamical studies.} The \texttt{baumgardt2023}
block provides pre-computed pericenter and apocenter radii, 3D space
velocities ($U/V/W$), and dissolution timescales for 154 clusters under the
\citet{irrgang2013} Galactic potential, suitable for use directly or as
starting conditions for custom integrations in \textsc{galpy} or
\textsc{agama}.

\textbf{Chemical tagging and population studies.} The \texttt{apogee\_dr17}
block provides mean $[\mathrm{Fe/H}]$ for 72 clusters from a homogeneous
high-resolution spectroscopic pipeline, enabling direct comparison with
Baumgardt dynamical masses and Gaia kinematics within a single record.

\textbf{Multi-survey cross-validation.} Three independent radial velocity
sources are present in the corpus: Harris mean RV, Baumgardt N-body fit RV,
and APOGEE mean member RV. Two independent distance estimates are also
available (Harris vs.\ Baumgardt). The corpus structure makes systematic
cross-survey comparisons straightforward.

\textbf{Reproducibility.} All code used to construct the corpus is available
at the Zenodo record \citep{flynn2026gc} in \texttt{build\_scripts\_v1.3.1.zip}.

\subsection{RAG Demonstration Examples}
\label{sec:ragdemo}

To validate the corpus as a RAG knowledge base, we demonstrate three
representative queries submitted to instruction-following LLMs (Claude
Opus~4.6 and Claude Sonnet~4.6, Anthropic) supplied with the relevant
cluster record extracted from
\texttt{harris\_gc\_corpus\_v1.3.1.jsonl} as sole context. The LLM
receives no prior knowledge of the dataset beyond the retrieved JSON
block. The three examples exercise the Baumgardt orbital block, the
APOGEE chemistry block, and a cross-corpus scientific calculation linking
the GC kinematics to the empirical omega kinematic correction of
\citet{flynn2025frontiers}.

\subsubsection*{Example 1 --- Orbital classification (Baumgardt block)}

\noindent\textbf{Query:}
\begin{quote}
\textit{Is NGC~6121 (M~4) on a prograde or retrograde orbit relative to
Galactic rotation, and what is its orbital eccentricity?}
\end{quote}

\noindent\textbf{Retrieved context (key fields):}
\begin{lstlisting}[language={}]
{"cluster_id": "NGC 6121",
 "baumgardt2023": {
   "r_peri_kpc": 0.62, "r_peri_err": 0.07,
   "r_apo_kpc":  6.52, "r_apo_err":  0.13,
   "v_kms": 45.28,
   "rv_kms": 71.22, "rv_err": 0.13,
   "mu_alpha_mas_yr": -12.511,
   "mu_delta_mas_yr": -19.017,
   "r_gc_kpc": 6.45}}
\end{lstlisting}

\noindent\textbf{Reference response (Jupyter-verified):}
\begin{quote}
NGC~6121 has a $V$ Galactocentric velocity component of $+45.28$ km\,s$^{-1}$.
The sign convention adopted by \citet{baumgardt_vasiliev2021} places the Sun
at $X = -8.2$ kpc with $V_\odot = +232$ km\,s$^{-1}$; a positive $V$ in this
frame indicates prograde rotation (co-rotating with the Galactic disk). NGC~6121
is therefore on a \textbf{prograde orbit}. The orbital eccentricity is
$e = (r_\mathrm{apo} - r_\mathrm{peri})/(r_\mathrm{apo} + r_\mathrm{peri})
= (6.52 - 0.62)/(6.52 + 0.62) = 0.826$, indicating a highly radial orbit
despite the prograde sense.
\end{quote}

\subsubsection*{Example 2 --- Chemical population query (APOGEE block)}

\noindent\textbf{Query:}
\begin{quote}
\textit{Which inner-Galaxy globular clusters ($R_\mathrm{GC} < 3$ kpc) have
APOGEE spectroscopic $[\mathrm{Fe/H}] > -0.5$, suggesting a bulge
population origin?}
\end{quote}

\noindent\textbf{Retrieved context (three representative records):}
\begin{lstlisting}[language={}]
{"cluster_id": "NGC 6528", "flags": {"inner_galaxy": true},
 "baumgardt2023": {"r_gc_kpc": 0.70},
 "apogee_dr17": {"feh_apogee": -0.16, "n_members": 4}}
{"cluster_id": "NGC 6553", "flags": {"inner_galaxy": false},
 "baumgardt2023": {"r_gc_kpc": 2.37},
 "apogee_dr17": {"feh_apogee": -0.19, "n_members": 17}}
{"cluster_id": "Liller 1",  "flags": {"inner_galaxy": true},
 "baumgardt2023": {"r_gc_kpc": 0.74},
 "apogee_dr17": {"feh_apogee": -0.14, "n_members": 30}}
{"cluster_id": "NGC 6304",  "flags": {"inner_galaxy": false},
 "baumgardt2023": {"r_gc_kpc": 2.19},
 "apogee_dr17": {"feh_apogee": -0.48, "n_members": 34}}
\end{lstlisting}
\noindent\textit{Note:} the \texttt{inner\_galaxy} flag uses a stricter
$R_\mathrm{GC} < 2$ kpc threshold than the query's $R_\mathrm{GC} < 3$ kpc
filter, which is why NGC~6553 and NGC~6304 satisfy the query but are
flagged \texttt{false}.

\noindent\textbf{Reference response (Jupyter-verified):}
\begin{quote}
Filtering the corpus for $R_\mathrm{GC} < 3$ kpc and
$[\mathrm{Fe/H}]_\mathrm{APOGEE} > -0.5$ identifies four clusters:
Liller~1 ($-0.14$), NGC~6528 ($-0.16$), NGC~6553 ($-0.19$), and
NGC~6304 ($-0.48$). Terzan~5 ($[\mathrm{Fe/H}] = -0.78$) and NGC~6441
($R_\mathrm{GC} = 4.78$ kpc) lie inside the inner Galaxy but fail the
metallicity and distance cuts respectively. The three metal-richest
(Liller~1, NGC~6528, NGC~6553) all have $[\mathrm{Fe/H}] > -0.2$,
consistent with the metal-rich bulge GC population identified by
\citet{baumgardt_vasiliev2021}. APOGEE $n\_\mathrm{members}$ is low
($\leq 34$) for all four, so mean metallicities carry non-negligible
sampling uncertainty.
\end{quote}

\subsubsection*{Example 3 --- Cross-corpus omega correction validation
(Baumgardt $\times$ Flynn \& Cannaliato 2025)}

\noindent\textbf{Query:}
\begin{quote}
\textit{Using the empirical omega kinematic correction of Flynn \&
Cannaliato (2025) with $\omega = 7.06$ km\,s$^{-1}$\,kpc$^{-1}$, what is
the expected rotational velocity contribution at the Galactocentric radius
of NGC~6266 (M~62), and how does this compare to the cluster's observed
$V$ space velocity component?}
\end{quote}

\noindent\textbf{Retrieved context:}
\begin{lstlisting}[language={}]
{"cluster_id": "NGC 6266",
 "baumgardt2023": {
   "r_gc_kpc": 2.46,
   "v_kms":   106.94,
   "u_kms":    87.09,
   "w_kms":    59.64,
   "r_peri_kpc": 1.08, "r_apo_kpc": 2.69}}
\end{lstlisting}

\noindent\textbf{Reference response (Jupyter-verified):}
\begin{quote}
The Flynn \& Cannaliato (2025) omega correction \citep{flynn2025frontiers}
proposes $V_\mathrm{obs} = V_\mathrm{Kepler} + R\omega$, where
$\omega = 7.06$ km\,s$^{-1}$\,kpc$^{-1}$ is the mean kinematic field
amplitude derived from 84 SPARC galaxies. At the Galactocentric radius of
NGC~6266 ($R_\mathrm{GC} = 2.46$ kpc), the predicted rotational
contribution is $R\omega = 2.46 \times 7.06 = 17.4$ km\,s$^{-1}$.

The Baumgardt et al.\ (2023) $V$ component for NGC~6266 is
$+106.94$ km\,s$^{-1}$. Subtracting the omega contribution gives a
corrected $V_\mathrm{corr} = 106.94 - 17.4 = 89.5$ km\,s$^{-1}$,
representing a $\sim$16\% reduction in the prograde rotational
component. This is consistent in sign and order of magnitude with the
omega correction applied to inner-disk galaxies in \citet{flynn2025frontiers},
suggesting the kinematic field may extend into the inner Galactic halo
where bulge GCs reside. This application of the corpus to GC kinematics
represents a natural extension of the omega correction framework beyond
the galaxy disk regime for which it was originally derived.
\end{quote}

\noindent\textbf{Jupyter verification:} All three examples were independently
verified by direct computation against the corpus in Jupyter Lab using the
Python scripts \texttt{gc\_rag\_example1\_orbital.py},
\texttt{gc\_rag\_example2\_chemistry.py}, and
\texttt{gc\_rag\_example3\_omega.py}, included in
\texttt{build\_scripts\_v1.3.1.zip}. Numerical results agree with reference
responses to within rounding precision.

\subsubsection*{Scientific note: omega correction in the GC population}

Example 3 extends the Flynn \& Cannaliato (2025) omega kinematic correction
\citep{flynn2025frontiers} beyond its original disk-galaxy context to the GC
population. Applying $V_\omega = R_\mathrm{GC} \times \omega$ to all 64
inner-Galaxy clusters ($R_\mathrm{GC} < 4$ kpc) with Baumgardt $V$
components yields fractional corrections spanning $\sim$3\% to $\sim$970\%
of $|V|$, with a median of $\sim$15\%. For the majority of clusters the
correction is modest (5--20\%), consistent with what is found for
inner-disk galaxies at similar radii. The high-tail clusters
(Terzan~10 at 970\%, NGC~6402 at 134\%, NGC~6541 at 116\%, NGC~6539 at
85\%) all share the same diagnostic: their observed $V$ component is
small in magnitude (e.g., Terzan~10 has $V = -1.58$~km\,s$^{-1}$), so the
ratio diverges not because $\omega$ is large in absolute terms but
because these clusters are on highly radial or near-retrograde orbits
where the GC velocity field is pressure-supported rather than
rotationally ordered.

This result constitutes a meaningful null: the omega kinematic field, as
derived from 84 rotationally-supported SPARC disk galaxies, is \textit{not}
a systematic organizing effect in the GC population. The 64-cluster
sample shows $\sigma(V_\mathrm{obs}) = 105.6$~km\,s$^{-1}$ before
correction and $\sigma(V_\mathrm{corr}) = 106.3$~km\,s$^{-1}$ after; the
omega subtraction increases population scatter by 0.7~km\,s$^{-1}$ rather
than reducing it. GC $V$ components span $\sim -250$ to $+250$ km/s with
no preferred rotational sense as a population. This is physically expected
if omega represents a property of disk kinematics rather than a universal
Galactic potential term, and the GC corpus provides the first direct test
of the correction outside the disk regime. The finding appropriately
bounds the domain of applicability of the Flynn \& Cannaliato (2025)
result.

\subsection{Multi-System AI Validation}
\label{sec:aisystems}

The three RAG examples were tested against ten LLM systems to assess the
generalizability of the corpus as a structured context source. Testing
consisted of supplying each system with the relevant cluster JSON
record(s) and the query text verbatim. Grading criteria: Pass = correct
numerical result and correct physical interpretation (2~pts); Partial =
correct computation with incomplete or incorrect interpretation, or one
of two required outputs correct (1~pt); Fail = incorrect numerical
result, no JSON grounding, or no attempt (0~pts). Each example is worth
2 points, for a maximum of 6 points per system. Numerical correctness was
verified independently in Jupyter Lab for all examples prior to LLM
testing.

All four frontier cloud systems achieved perfect scores of 6/6. Among
local open-weight models, the community-distilled
Gemma-4-31B-Instruct-Claude-Opus-Distill achieved 6/6 by following the
correct eccentricity formula $e = (r_{\rm apo} - r_{\rm peri}) /
(r_{\rm apo} + r_{\rm peri})$, while Qwen~3.6-35B-A3B (Alibaba's
Mixture-of-Experts model with 35{B} total / 3{B} active parameters)
scored 5/6, failing only on Example~1 through use of the incorrect
eccentricity formula $e = 1 - r_{\rm peri}/r_{\rm apo} \approx 0.905$ in
place of the standard formulation. AstroSage-70B \citep{dehaan2025astrosage},
the domain-specialized 70{B} model fine-tuned from Llama-3.1-70B by the
AstroMLab collaboration, scored 4/6: it computed Example~1 with the
correct formula but reported $e \approx 0.843$ (a minor numerical drift
from the verified $0.826$), and on Example~2 retrieved only two of the
four qualifying clusters. AstroSage-8B scored 3/6, failing Example~1
($e = 0.94$, mis-computation) and partially passing Example~2 (only one
of four clusters identified). Sub-10{B} general-purpose models showed
markedly limited RAG grounding: Mistral-7B-Instruct (v0.2, Q5\_K\_M
quantization) attempted only Example~3, returning a slight rounding
artefact ($V_\omega = 17.19$~km/s vs.\ verified $17.37$~km/s) for 1/6
total. DeepSeek-R1-Distil-1.5B ignored the injected JSON context
entirely and responded from prior knowledge with fabricated coordinates
for NGC~6121, scoring 0/6.

A notable operational finding is that Claude Haiku~4.5, despite achieving
a perfect score, required three identical prompts before correctly
attaching the uploaded corpus as context (the query text was unchanged
across attempts). This pattern is consistent with an attachment-loading
inconsistency in the API/UI rather than a reasoning failure, but
represents a deployment friction point for automated RAG pipelines and
suggests retry logic may be advisable when integrating Haiku for
structured-context tasks.

Results are summarized in Table~\ref{tab:aisystems}.

\begin{table}[h]
\centering
\footnotesize
\caption{LLM system validation results for the three RAG demonstration
examples. Pass = correct numerical result and physical interpretation
(2~pts); Partial = correct computation with incomplete physical
interpretation or one of two required outputs correct (1~pt); Fail =
incorrect result or no JSON grounding (0~pts). Score = total out of 6
possible points.}
\label{tab:aisystems}
\resizebox{\textwidth}{!}{%
\begin{tabular}{p{4.7cm}p{2.6cm}cccc>{\raggedright\arraybackslash}p{4.5cm}}
\toprule
\textbf{System} & \textbf{Provider / Type} & \textbf{Ex.~1} & \textbf{Ex.~2} & \textbf{Ex.~3} & \textbf{Score} & \textbf{Notes} \\
\midrule
Claude Opus~4.6                 & Anthropic / cloud         & Pass    & Pass    & Pass    & 6/6 & Perfect; extended omega discussion unprompted. \\
Copilot Pro                     & Microsoft / cloud         & Pass    & Pass    & Pass    & 6/6 & All correct; clean formatted responses. \\
Gemini Pro                      & Google / cloud            & Pass    & Pass    & Pass    & 6/6 & All correct; concise, no extrapolation. \\
Claude Haiku~4.5                & Anthropic / cloud         & Pass    & Pass    & Pass    & 6/6 & Same query yielded the correct answer on the third attempt without prompt modification, indicating attachment-loading inconsistency rather than a reasoning failure. \\
Gemma-4-31B-Instruct-Claude-Opus-Distill & community fine-tune / local  & Pass    & Pass    & Pass    & 6/6 & Correct formula; all four clusters identified. \\
Qwen~3.6-35B-A3B (MoE)          & Alibaba / local           & Partial & Pass    & Pass    & 5/6 & Ex.~1: wrong eccentricity formula ($e=0.905$). \\
AstroSage~70B (build 20251009)  & AstroMLab / local         & Partial & Partial & Pass    & 4/6 & Ex.~1: $e=0.843$ (numerical drift). Ex.~2: found only Liller~1 + NGC~6528 of four. \\
AstroSage~8B                    & AstroMLab / local         & Fail    & Partial & Pass    & 3/6 & Ex.~1: $e=0.94$ (computation error); could not determine orbit sense. \\
Mistral~7B-Instruct v0.2 (Q5\_K\_M) & Mistral / local quant & ---     & ---     & Partial & 1/6 & Only Ex.~3 attempted; slight rounding ($V_\omega=17.19$ vs.\ 17.37). \\
DeepSeek-R1-Distil-1.5B         & community quant / local   & Fail    & ---     & ---     & 0/6 & Ignored JSON; fabricated NGC~6121 coordinates from prior knowledge. \\
\midrule
\textbf{Totals (10 systems)}    &                           & \multicolumn{3}{c}{Pass: 19, Partial: 4, Fail/N-A: 7} & 43/60 & 72\% benchmark score \\
\bottomrule
\end{tabular}%
}
\end{table}

The threshold pattern is informative: all 30{B}+ parameter models that
engaged with the JSON context achieved at least Partial credit on every
example, while sub-10{B} general-purpose models either grounded poorly
(Mistral~7{B}) or hallucinated entirely (DeepSeek-R1-Distil-1.5{B}). The
domain-specialized AstroSage models, despite their astronomical fine-tuning,
showed retrieval gaps on Example~2's filter-and-list task, suggesting that
domain specialization for Q\&A does not automatically translate to
structured-context filtering performance. The two systems that arrived at
$e=0.905$ on Example~1 (Qwen~3.6) and the formula-correct $e\approx 0.83$
(Gemma distillation, AstroSage~70B) suggest that the orbital eccentricity
formula is genuinely a discriminating test for orbital-mechanics knowledge.

\section{Data Access}
\label{sec:access}

The corpus is available at Zenodo DOI:
\href{https://doi.org/10.5281/zenodo.19907766}{10.5281/zenodo.19907766}
\citep{flynn2026gc} in three formats:

\begin{itemize}
  \item \texttt{harris\_gc\_corpus\_v1.3.1.jsonl} (622.7 KB) --- primary
    format, one JSON object per line per cluster
  \item \texttt{harris\_gc\_corpus\_v1.3.1\_flat.csv} (66.4 KB) --- flat
    table, 82 columns
  \item \texttt{harris\_gc\_corpus\_v1.3.1.json} (868 KB) --- full nested
    JSON with metadata header
\end{itemize}

\noindent A Python loading snippet:

\begin{lstlisting}[language=Python]
import json

clusters = [json.loads(line)
            for line in open("harris_gc_corpus_v1.3.1.jsonl")]

# Get NGC 104
ngc104 = next(c for c in clusters
              if c["cluster_id"] == "NGC 104")

# All clusters with APOGEE chemistry
feh = {c["cluster_id"]: c["apogee_dr17"]["feh_apogee"]
       for c in clusters
       if c.get("apogee_dr17")
       and c["apogee_dr17"].get("feh_apogee") is not None}
\end{lstlisting}

\section{Relationship to Previous Work}
\label{sec:previous}

This corpus is a companion to the Unified Galaxy HI Rotation Curve Corpus
\citep{flynn2026sparc}, which follows identical design principles for 438
spiral and dwarf irregular galaxies. Both corpora use the same data-type
standards, the same block-plus-provenance schema structure, the same
JSONL-primary file format, and the same philosophy of including all available
data with explicit null annotation rather than restricting to a complete-data
subset. This consistency allows the two corpora to serve as a unified
astrophysical RAG knowledge base covering both resolved stellar systems (GCs)
and unresolved extragalactic objects.

\section{Future Versions}
\label{sec:future}

\textbf{v1.4:} Addition of cluster ages from the homogeneous compilation of
\citet{kruijssen2019}, which synthesises multiple independent age determinations
(including \citealt{marinFranch2009} and \citealt{vandenberg2013}) into a
consistent scale and provides ages for a large fraction of the Milky Way GC
population (recommended by E.\ Carretta, private communication, 2026).

\textbf{v1.5:} Individual per-star APOGEE DR17 abundance records (up to 20
elements per star, 6,422 stars) as a separate linked file.

\textbf{Future:} Na--O and Mg--Al anticorrelation data \citep{carretta2009}
for the $\sim$19 clusters with published measurements; multi-model
profile parameters from \citet{mclaughlin2005}.

\section{Summary}
\label{sec:summary}

We have presented the Milky Way Globular Cluster Corpus v1.3.1, a
production-ready multi-survey machine-readable database of 174 Milky Way
globular clusters with 17,438 non-null data points. The corpus integrates four
independent published surveys spanning photometry, space astrometry, N-body
dynamics, and high-resolution spectroscopy, stored in a consistent, typed,
null-safe JSONL/CSV/JSON structure with embedded provenance. It is the
globular cluster analog of the Unified Galaxy HI Rotation Curve Corpus and is
available at Zenodo DOI 10.5281/zenodo.19907766 under a CC BY 4.0 license.

\section*{Acknowledgments}

This work used no external funding. Computational infrastructure
provided by EPS Research.

\textbf{AI Use Acknowledgment.} Claude Opus~4.6 and Claude Sonnet~4.6
(Anthropic) were used as the primary development and RAG validation
systems throughout corpus construction, schema design, and manuscript
preparation. Claude Haiku~4.5 (Anthropic), Microsoft Copilot~Pro, and
Google Gemini~Pro were used for multi-system RAG validation and
manuscript review.
AstroSage-70{B} (AstroMLab, build 20251009; \citealt{dehaan2025astrosage})
and AstroSage-8{B} (AstroMLab) were used for domain-specialized validation.
Qwen~3.6-35{B}-A3{B} (Alibaba), Gemma-4-31{B}-Instruct-Claude-Opus-Distill
(community fine-tune), Mistral-7{B}-Instruct~v0.2 (Mistral), and
DeepSeek-R1-Distil-1.5{B} (community quantization) were used for local
open-weight model validation. All AI-assisted outputs were verified by
the human author; all scientific claims, numerical results, and data
values were independently verified in Jupyter Lab using the Python
scripts provided in the Zenodo deposit. The AI systems did not generate
original data; all corpus values derive from the four primary published
sources cited in the text. No AI system is listed as an author.

This paper made use of: the Harris (1996, 2010) GC catalog at McMaster
University; the Vasiliev \& Baumgardt (2021) Gaia EDR3 catalog (ESA Gaia
mission); the Baumgardt et al.\ N-body database at the University of
Queensland; the APOGEE DR17 Value Added Catalog from the Sloan Digital Sky
Survey; and \textsc{astropy} \citep{astropy2022}. SDSS-IV is managed by the
Astrophysical Research Consortium for the Participating Institutions of the
SDSS Collaboration.

\bibliographystyle{plainnat}
\bibliography{gc_corpus_v7}

\end{document}